\documentclass[journal=jpclcd,manuscript=letter]{achemso}
\usepackage{amsmath,amssymb}
\usepackage{float}
\usepackage{upgreek}
\usepackage{color}
\usepackage{array}
\usepackage[version=3]{mhchem} 

\author{Manish Kumar}
\email{manish.kumar@physics.iitd.ac.in[MK]}
\author{Arunima Singh, Deepika Gill}
\author{Saswata Bhattacharya}
\email{saswata@physics.iitd.ac.in[SB]}
\phone{+91-11-2659 1359}
\affiliation[Indian Institute of Technology Delhi]
{Department of Physics, Indian Institute of Technology Delhi, New Delhi, India}
\title[An \textsf{achemso} demo]
{Optoelectronic Properties of Chalcogenide Perovskites by Many-Body Perturbation Theory}
\begin{document}
\begin{abstract}
Chalcogenide perovskites have emerged as non-toxic and stable photovoltaic materials, acting as an alternative to lead halide hybrid perovskites having similar optoelectronic properties. In the present work, we report the electronic and optical properties of chalcogenide perovskites AZrS$_3$ (A=Ca, Sr, Ba) by using the density functional theory (DFT) and many-body perturbation theory (MBPT viz. G$_0$W$_0$ and BSE). This study includes excitonic analysis for the aforementioned systems. The exciton binding energy (E$_\textrm{B}$) is found to be larger than that of the halide perovskites, as the ionic contribution to dielectric screening is negligible in the former. We also observe a more stable charge-separated polaronic state as compared to that of the bound exciton. Finally, on the basis of direct gap and absorption coefficient, the estimated spectroscopic limited maximum efficiency (SLME) of the solar cells is large and suggests the applicability of these perovskites in photovoltaics.
  \begin{tocentry}
  \begin{figure}[H]%
  	\includegraphics[width=1.0\columnwidth,clip]{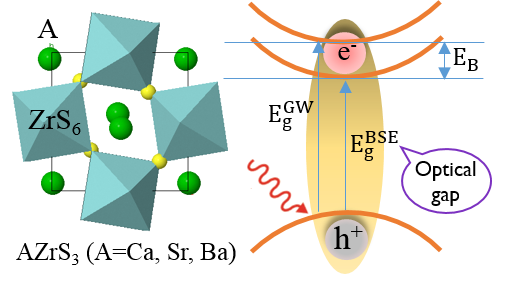}
  \end{figure}	
  \end{tocentry}
\end{abstract}
Inorganic-organic (IO) hybrid halide perovskites have emerged as an efficient compound semiconductors alternative to conventional materials used in photovoltaics\cite{doi:10.1021/ja809598r,Lee643,doi:10.1021/acs.chemrev.8b00539,nrel}. The power conversion efficiency (PCE) of solar cell based on IO hybrid perovskites has increased from 3.8\% to 25.5\% in the last decade\cite{doi:10.1021/ja809598r,nrel}. Nevertheless, the concerns regarding long term stability and toxicity of lead restrict their commercialization\cite{babayigit2016toxicity,pooja-prb}. In pursuit of alternative materials having similar kind of optoelectronic properties, chalcogenide perovskites have been investigated in last few years. Sun \textit{et al.}\cite{doi:10.1021/nl504046x} first theoretically reported  the optoelectronic properties of chalcogenide perovskites, which can be utilized in solar cells. Subsequently, other theoretical studies have also characterized different chalcogenide perovskites for high solar cell efficiency and photoelectrochemical water splitting\cite{doi:10.1021/acs.chemmater.5b04213,C7EE02702H,C9TA03116B,ma13040978,PhysRevMaterials.3.101601}. Many chalcogenide perovskites such as AZrS$_3$ (A=Ca, Sr, Ba), have been synthesized experimentally as well\cite{https://doi.org/10.1107/S056774088000845X,LEE20051049,doi:10.1021/acs.chemmater.5b04213,PERERA2016129,https://doi.org/10.1002/adma.201604733,doi:10.1021/acsaem.9b02428,doi:10.1021/acsomega.0c00740,D0NR08078K}, which are stable and consist of earth-abundant non-toxic elements. These chalcogenide perovskites have high optical absorption, optimal photoluminescence and good charge carrier mobility, which suggest the possibility of their usage in various optoelectronic devices\cite{https://doi.org/10.1002/adma.201604733,doi:10.1021/jacs.8b13622,doi:10.1021/acs.chemmater.8b04178,https://doi.org/10.1002/solr.201900555}.

Zr-based chalcogenide perovskites contain the d-orbital character, wherein the 4d states are less localized than 3d, resulting in large absorption coefficient and small effective mass of the charge carriers in these compounds\cite{doi:10.1021/nl504046x}. Therefore, many experimental and theoretical studies have been performed on Zr-based chalcogenide perovskites (AZrS$_3$, where A is alkaline earth metal).

Note that, the charge separation in solar cell gets hugely influenced by formation of excitons. Therefore, the operation mechanism of a solar cell highly depends on it as these excitons thermally dissociate into free electrons and holes, giving rise to the required free‐charge transport. However, until date, due to huge computational cost, any detailed study with adequate accuracy of the excitonic properties is not very well known. Therefore, it is of profound interest to employ advanced theoretical methodologies for accurate understanding of the excitonic properties that will sufficiently correlate with the experimental studies to disentangle the scientific insights of excitons. Despite several theoretical studies on the chalcogenide perovskites, investigation of the optical properties using the excited-state methods remain unexplored. In view of this, presumably for the first time, we have reported the excitonic properties of the chalcogenide perovskites.

In this Letter, we have done a systematic study of electronic and optical properties of chalcogenide perovskites AZrS$_3$ (A=Ca, Sr, Ba) using ground- and excited-state methods. First, we have employed density functional theory (DFT)~\cite{PhysRev.136.B864,PhysRev.140.A1133} with semi-local PBE~\cite{PhysRevLett.77.3865} exchange-correlation ($\epsilon_\textrm{xc}$) functional to optimize the crystal structures. In order to study electronic structure, we have calculated atom-projected electronic partial density of states (pDOS) using hybrid $\epsilon_\textrm{xc}$ functional HSE06~\cite{doi:10.1063/1.2404663}. Subsequently, we have determined the optical properties using excited-state method viz. many-body perturbation theory (MBPT). The Bethe-Salpeter equation (BSE)~\cite{PhysRevLett.80.4510,PhysRevLett.81.2312} has been solved to get the electronic contribution to dielectric function on top of single-shot G$_0$W$_0$@PBE~\cite{PhysRev.139.A796,PhysRevLett.55.1418}. Further to investigate the ionic contribution to dielectric function, density functional perturbation theory (DFPT) has been used. Finally, using the quasiparticle (QP) band gap and optical properties, the maximum theoretical photoconversion efficiency has been determined by calculating the spectroscopic limited maximum efficiency (SLME)~\cite{PhysRevLett.108.068701} metric.

\begin{figure}[h]
	\centering
	\includegraphics[width=0.8\textwidth]{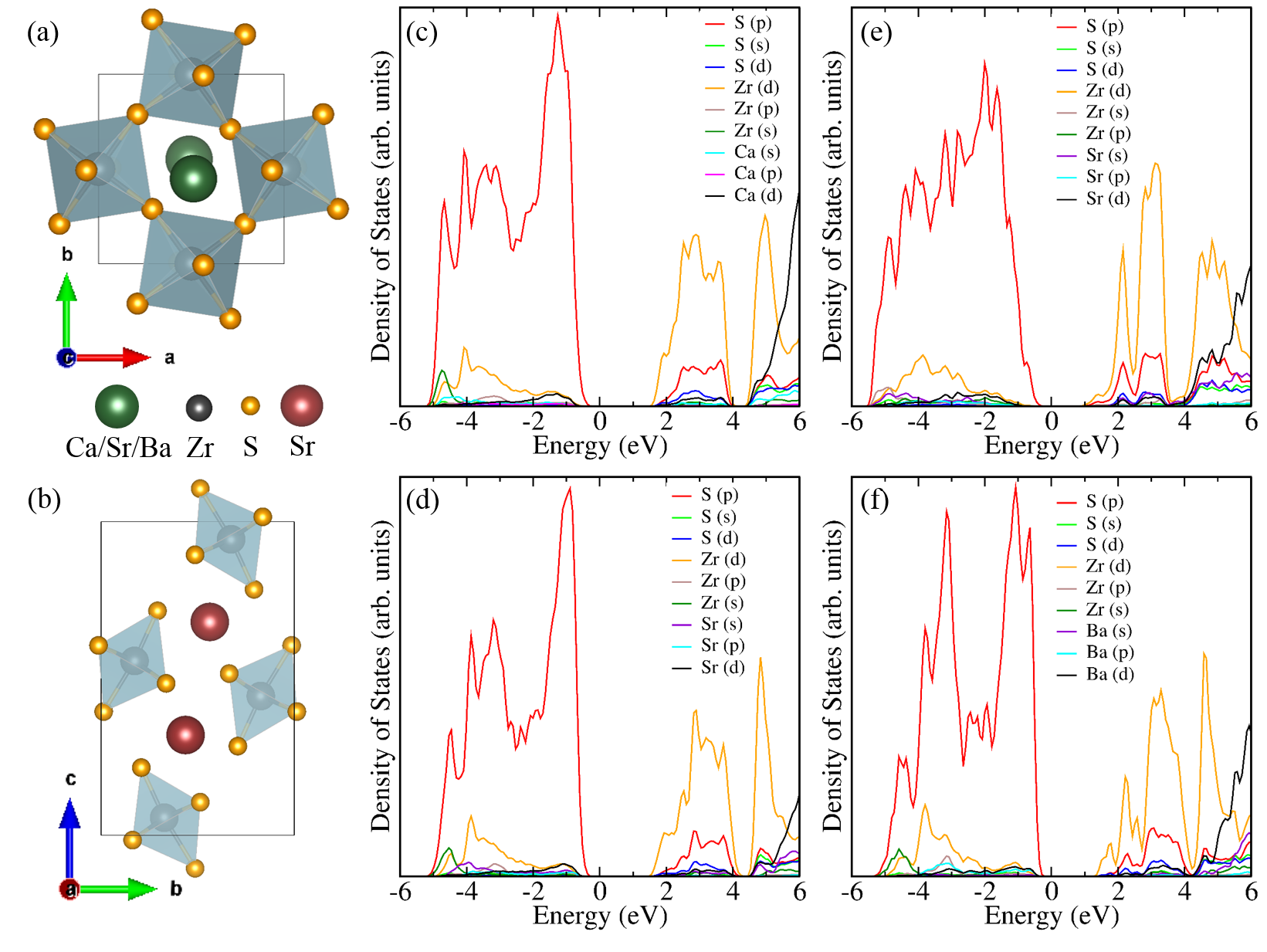}
	\caption{Schematic crystal structure of orthorhombic (a) AZrS$_3$ (A=Ca, Sr, Ba) in distorted phase and (b) $\alpha-$SrZrS$_3$ in needle-like phase. Electronic partial density of states (pDOS) of (c) CaZrS$_3$, (d) $\beta-$SrZrS$_3$, (e) $\alpha-$SrZrS$_3$, and (f) BaZrS$_3$ using HSE06 $\epsilon_\textrm{xc}$ functional.}
	\label{fig_dos}
\end{figure}

Here, we have considered the distorted orthorhombic phase of chalcogenide perovskites AZrS$_3$ (A=Ca, Sr, Ba) having the space group $Pnma$~\cite{https://doi.org/10.1107/S056774088000845X} (see Figure \ref{fig_dos}a). In addition, we have considered the needle-like phase of SrZrS$_3$ (see Figure \ref{fig_dos}b), since the same has been found in two crystallized phases ($\alpha-$phase, which is needle-like and $\beta$-phase, which is distorted-perovskite phase)\cite{LEE20051049}. The lattice parameters of the optimized structures calculated using PBE $\epsilon_\textrm{xc}$ functional are given in Table \ref{tbl:1}. These are in close agreement with previous experimental results\cite{https://doi.org/10.1107/S056774088000845X,LEE20051049}.
\begin{table}
	\caption{Calculated lattice parameters of AZrS$_3$ (A=Ca, Sr, Ba) perovskites. The experimental values are provided in brackets. For distorted perovskites viz. CaZrS$_3$, $\beta-$SrZrS$_3$, and BaZrS$_3$, the experimental values are from Ref. \cite{https://doi.org/10.1107/S056774088000845X}. For $\alpha-$SrZrS$_3$, experimental values are from Ref. \cite{LEE20051049}}
	\label{tbl:1}
	\centering
	\setlength\extrarowheight{+4pt}
	\begin{tabular}[c]{cccc}
		\hline
		Configurations & a (\AA) & b (\AA)&  c (\AA)\\
		\hline
		CaZrS$_3$ & 6.56 (6.54)& 7.06 (7.03)& 9.63 (9.59)\\
		$\alpha-$SrZrS$_3$ &3.84 (3.83)& 8.63 (8.53)&13.99 (13.92)\\
		$\beta-$SrZrS$_3$ &6.78 (6.74)&7.16 (7.11)&9.82 (9.77)\\
		BaZrS$_3$ &7.03 (7.03)&7.16 (7.06)&10.01 (9.98)\\
		\hline
	\end{tabular}
\end{table}
We have calculated the electronic pDOS of the aforementioned configurations using HSE06 $\epsilon_\textrm{xc}$ functional as shown in Figure \ref{fig_dos}[c-f]. The valence band maximum (VBM) is mostly contributed by S 3p-orbitals, whereas the conduction band minimum (CBm) is mainly from Zr 4d-orbitals. Rest of the orbitals have small contribution at VBM and CBm. As we change the cation from Ca to Ba, the peaks in pDOS become more narrow and sharper at the CBm (see Figure \ref{fig_dos}c, \ref{fig_dos}d and \ref{fig_dos}f), which indicates that electronic nonradiative lifetime becomes shorter as we go down the group. This is due to the fact that narrow peaks in the pDOS signify large number of carrier relaxation path, and hence, shorter carrier lifetimes~\cite{doi:10.1021/acsenergylett.9b02593,doi:10.1021/acs.nanolett.5b00109}. $\alpha-$SrZrS$_3$ has the narrowest and sharpest peak in the pDOS at CBm (see Figure \ref{fig_dos}e). Further, with the change in A-cation species, there is shift in CBm, which has altered the band gap. Moreover, as we go down the group from Ca to Ba, the bands become more dispersive at the CBm (see Figure S1). Hence, the effective mass of the electron decreases (see Table \ref{tbl:2}). Contrastingly, the effective mass of hole is not affected much. We find $\alpha-$SrZrS$_3$ to have the smallest electron and hole effective masses. Therefore, these perovskites are expected to have better charge carrier transport as indicated by the smaller values of effective masses.

\begin{table}
	\caption{Effective mass of electron, hole and reduced mass (in terms of free-electron mass m$_\textrm{e}$) of chalcogenide perovskites along $\Gamma-$Z high symmetry path}
	\label{tbl:2}
	\centering
	\setlength\extrarowheight{+4pt}
	\begin{tabular}[c]{cccc}
		\hline
		Configurations & m$_\textrm{e}^*$ & m$_\textrm{h}^*$ &  $\mu$\\
		\hline
		CaZrS$_3$ & 0.503& 0.588& 0.271\\
		$\alpha-$SrZrS$_3$ &0.323& 0.545&0.203\\
		$\beta-$SrZrS$_3$ &0.440&0.580&0.250\\
		BaZrS$_3$ &0.411&0.587&0.242\\
		\hline
	\end{tabular}
\end{table}
\begin{table}
	\caption{Band gap (in eV) of chalcogenide perovskites}
	\label{tbl:3}
	\centering
	\setlength\extrarowheight{+4pt}
	\begin{tabular}{ccccc}
		\hline
		Configurations &PBE& HSE06 & G$_0$W$_0$@PBE &  Experimental\\
		\hline
		CaZrS$_3$ &1.24& 2.04& 2.29& 1.90\cite{PERERA2016129}\\
		$\alpha-$SrZrS$_3$ &0.60 &1.40&1.60&1.52\cite{https://doi.org/10.1002/adma.201604733}\\
		$\beta-$SrZrS$_3$ &1.22 &2.05&2.32&2.05\cite{https://doi.org/10.1002/adma.201604733}\\
		BaZrS$_3$ &1.06 &1.87&2.10&1.83\cite{https://doi.org/10.1002/adma.201604733}\\
		\hline
	\end{tabular}
\end{table}
\begin{figure}[h]
	\centering
	\includegraphics[width=0.4\textwidth]{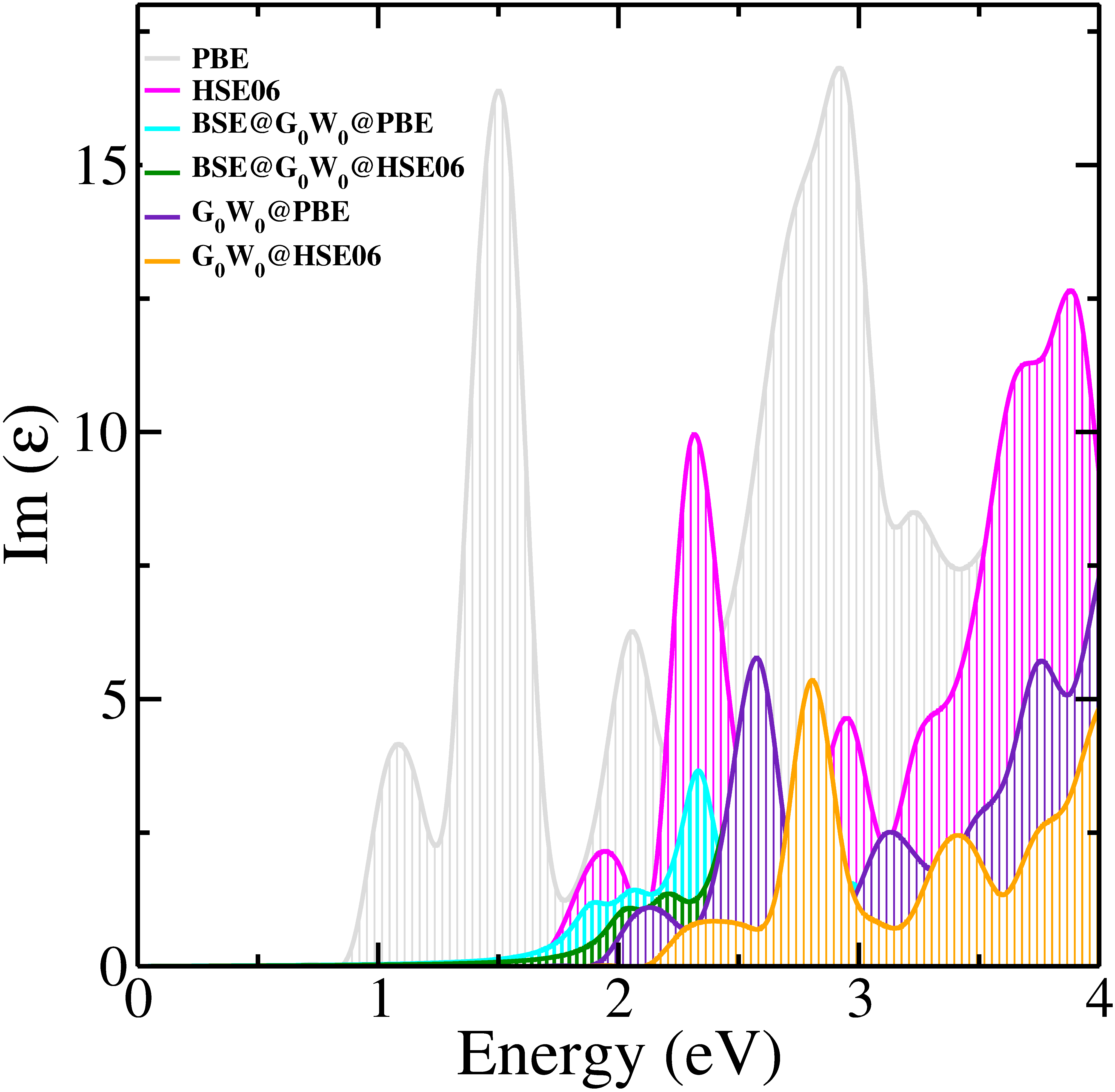}
	\caption{Imaginary [Im $(\upvarepsilon)$] part of the dielectric function for BaZrS$_3$ with light polarization perpendicular to c-axis ($\upvarepsilon_\textrm{xx}$), obtained using different level of theories viz. PBE, HSE06, G$_0$W$_0$@PBE, G$_0$W$_0$@HSE06, BSE@G$_0$W$_0$@PBE and BSE@G$_0$W$_0$@HSE06. First peak corresponds to the band gap of BaZrS$_3$.}
	\label{fig_valid}
\end{figure}
\begin{figure}[h]
	\centering
	\includegraphics[width=0.8\textwidth]{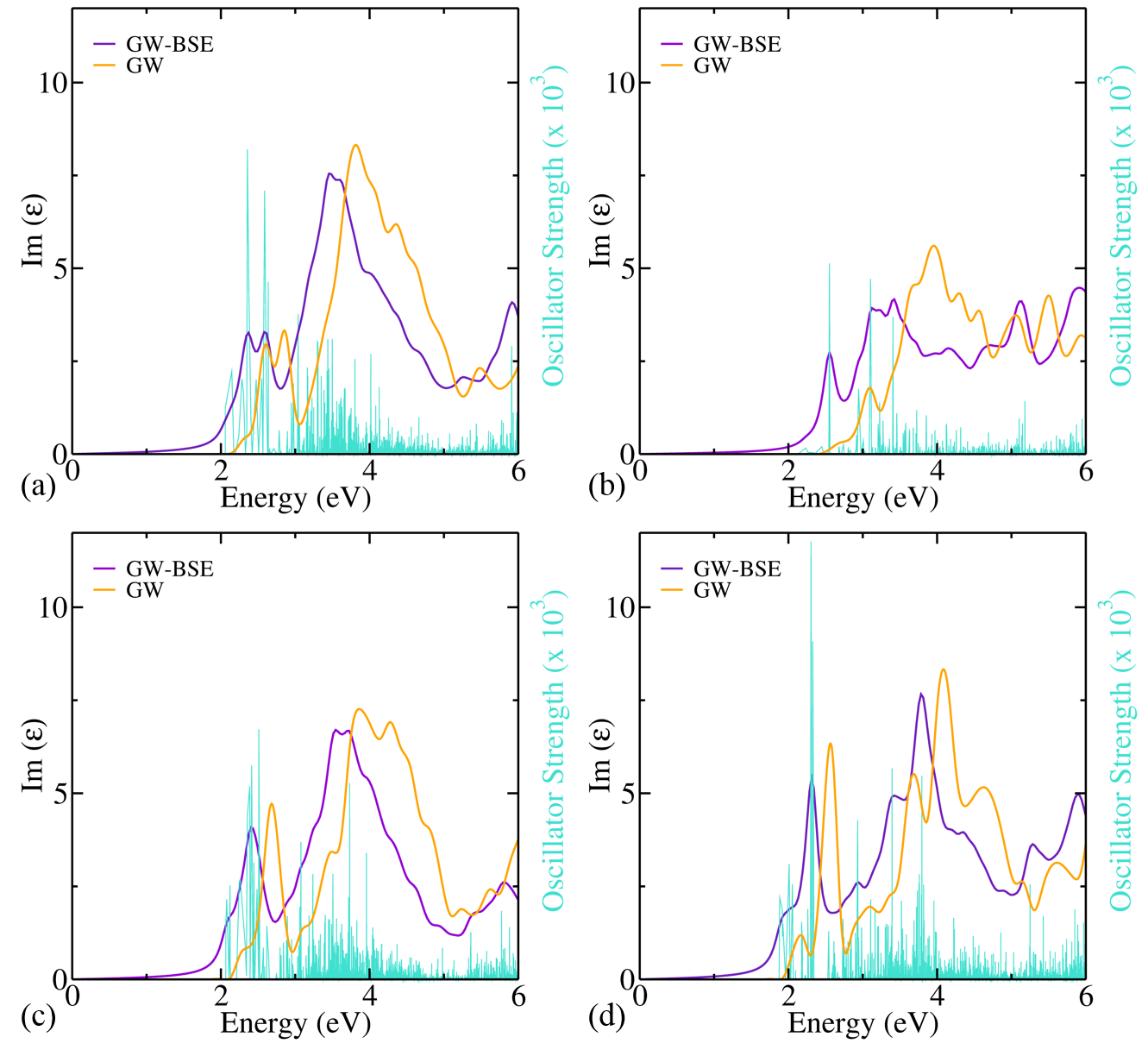}
	\caption{Spatially averaged imaginary [Im $(\upvarepsilon)$] part of the dielectric function for (a) CaZrS$_3$, (b) $\alpha-$SrZrS$_3$, (c) $\beta-$SrZrS$_3$, and (d) BaZrS$_3$ obtained using G$_0$W$_0$@PBE and BSE@G$_0$W$_0$@PBE. Peaks with turquoise color represent the oscillator strength.}
	\label{fig_optical}
\end{figure}
All the considered chalcogenide perovskites exhibit direct band gap at $\Gamma$ high symmetry point, which is the desired property for effective absorption. The band gaps are underestimated by semi-local $\epsilon_\textrm{xc}$ functional PBE (see Table \ref{tbl:3}), which is due to the well known self-interaction error. The hybrid $\epsilon_\textrm{xc}$ functional HSE06 corrects the band gaps that are well in agreement with the experimental values (see Table \ref{tbl:3}). Therefore, the electronic structure can be well described by the HSE06. However, the latter is less accurate in predicting the optical features of the systems\cite{PhysRevLett.107.216806}. Therefore, the MBPT based GW-BSE method has been used to compute the optical response, which explicitly considers the electron-hole interaction\cite{doi:10.1021/acs.jpcc.7b07473}. Since the single shot GW (G$_0$W$_0$) calculation depends on its starting point, we have validated it by calculating the optical response of BaZrS$_3$, i.e., the imaginary part of the complex dielectric function (see Figure \ref{fig_valid}). The first peak represents the optical transition corresponding to the band gap. The peak position, which is underestimated by PBE (1.06 eV), is improved by G$_0$W$_0$ on top of both the PBE and HSE06, which are at 2.10 and 2.32 eV, respectively (see Figure \ref{fig_valid}). Notably, the QP gaps computed using G$_0$W$_0$ are overestimated in comparison to experimental band gap, since it does not take into account the exciton binding energy. Further, the gaps are improved by solving the BSE. The peak positions obtained using BSE@G$_0$W$_0$@PBE and BSE@G$_0$W$_0$@HSE06 are at 1.88 and 2.02 eV, respectively (see Figure \ref{fig_valid}). The former is in close agreement with the experimental band gap, whereas the latter is overestimated. Hence, the PBE is accurate than the HSE06 as a starting point for calculating the optical properties using MBPT approach. The QP gaps of AZrS$_3$ (A=Ca, Sr, Ba) calculated using G$_0$W$_0$@PBE are provided in Table \ref{tbl:3}. Figure \ref{fig_optical} shows the imaginary part of the dielectric function [Im ($\upvarepsilon$)] and oscillator strength calculated using BSE@G$_0$W$_0$@PBE. In the same figure, Im ($\upvarepsilon$) calculated using G$_0$W$_0$@PBE is also shown. The exciton binding energy (E$_\textrm{B}$) can be computed from this figure, as the E$_\textrm{B}$ is the difference between QP band gap (G$_0$W$_0$@PBE peak position) and optical band gap (BSE@G$_0$W$_0$@PBE peak position). Hence, from Figure \ref{fig_optical}, the E$_\textrm{B}$ of the first bright exciton for CaZrS$_3$, $\alpha-$SrZrS$_3$, $\beta-$SrZrS$_3$ and BaZrS$_3$ are 0.23, 0.54, 0.25 and 0.21 eV, respectively. As per the BSE eigenvalue analysis, we have found that a dark exciton (optically inactive) also exists in case of BaZrS$_3$ below the bright exciton. Moreover, several dark excitons exist in case of $\alpha-$SrZrS$_3$. The E$_\textrm{B}$ for the lowest energetic dark exciton is 1.53 and 0.22 eV for $\alpha-$SrZrS$_3$ and BaZrS$_3$, respectively. Furthermore, the oscillator strength for all the considered perovskites is mainly distributed within the spectral window of 2-4 eV and matches well with the excitonic peak positions (see Figure \ref{fig_optical}). It signifies the high recombination between electron and hole in the considered energy range. Moreover, using the exciton binding energy, band gap, dielectric function and reduced mass, several excitonic parameters can be determined~\cite{basera2020capturing} such as excitonic temperature (T$_\textrm{exc}$) and radius (r$_\textrm{exc}$) given in Table \ref{Table1}. The lifetime ($\tau$) is inversely related to the probability of wavefunction ($|\phi_\textrm{n}(0)|^2$) for electron-hole pair at zero separation (for further details of excitonic parameters relations, see Section VIII in SI). 
Therefore, the $\tau$ for the considered perovskites are in the order: $\alpha$-$\textrm{SrZrS}_3>\textrm{BaZrS}_3>\textrm{CaZrS}_3>\beta$-$\textrm{SrZrS}_3$.
\begin{table}
	\caption{Excitonic parameters for chalcogenide perovskites} 
	\begin{center}
		\begin{tabular}[c]{ccccc} \hline
			Excitonic parameters  & CaZrS$_3$ & $\alpha$-SrZrS$_3$ & $\beta$-SrZrS$_3$  &  BaZrS$_3$ \\ \hline
			E$_\textrm{B}$ (eV)  & 0.23 & 0.54 & 0.25 & 0.21  \\ 
			T$_\textrm{exc}$ (K)  & 2669 & 6267 & 2901  & 2437  \\ 
			r$_\textrm{exc}$ (nm) & 0.79 & 0.93 & 0.73 & 0.92  \\ 
			$|\phi_\textrm{n}(0)|^2 (10^{27}$m$^{-3})$  & 0.65 & 0.40 & 0.82 &  0.41  \\
			\hline
		\end{tabular}
		\label{Table1}
	\end{center}
\end{table}

\begin{figure}[h]
	\centering
	\includegraphics[width=0.8\textwidth]{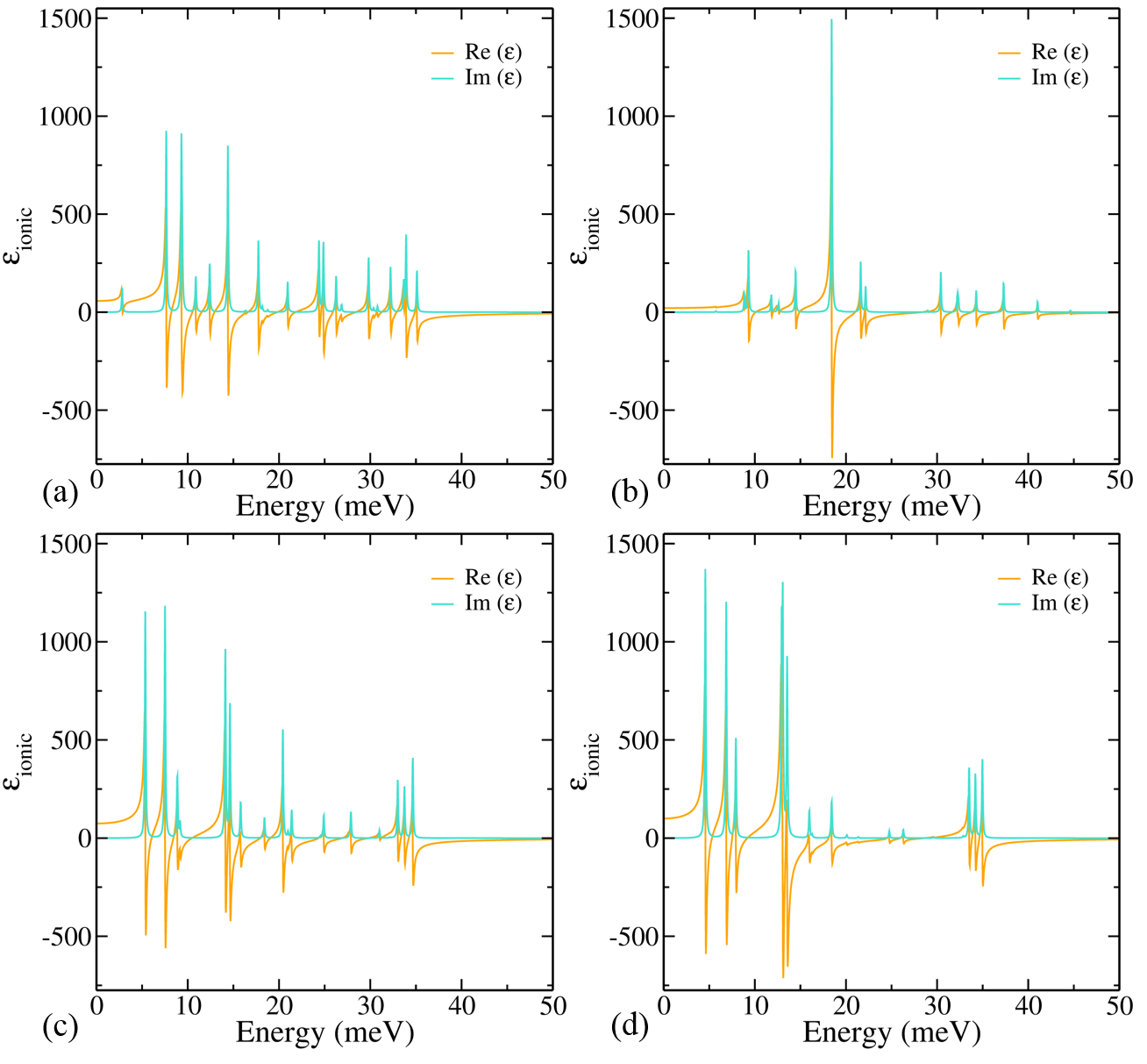}
	\caption{Ionic contribution to dielectric function for (a) CaZrS$_3$, (b) $\alpha-$SrZrS$_3$, (c) $\beta-$SrZrS$_3$, and (d) BaZrS$_3$ obtained using DFPT.}
	\label{fig_ionic}
\end{figure}

The high E$_\textrm{B}$ in comparison to halide perovskite\cite{doi:10.1021/acs.jpclett.8b02811} can be understood from the ionic contribution to dielectric screening. It has been recently shown that if the E$_\textrm{B}$ calculated using vertical transition is much greater than the energy of the longitudinal optical phonon mode ($\omega_\textrm{LO}$), then the ionic contribution to dielectric screening is negligible and hence, does not alter the E$_\textrm{B}$~\cite{bokdam2016role}. In case of chalcogenide perovskites, E$_\textrm{B} \gg \hbar\omega_\textrm{LO}$, which can be seen from the ionic contribution to dielectric function (see Figure \ref{fig_ionic}). Therefore, the lowering of E$_\textrm{B}$ by the ionic screening can be excluded. Further, by employing Wannier-Mott approach as well, we have calculated the E$_\textrm{B}$. According to this model, the E$_\textrm{B}$ is related to reduced mass of the charge carriers ($\mu$) and effective dielectric constant ($\varepsilon_{eff}$) as follows:
\begin{equation}
\textrm{E}_\textrm{B}=\frac{\mu}{\varepsilon_{eff}^2}\textrm{R}_\infty
\label{eq}
\end{equation}
\begin{table}
	\caption{Upper and lower bounds on exciton binding energy E$_\textrm{B}$ for chalcogenide perovskites}
	\label{tbl:bound}
	\centering
	\setlength\extrarowheight{+4pt}
	\begin{tabular}[c]{ccc}
		\hline
		Configurations & Upper bound (eV)& Lower bound (meV)\\
		\hline
		CaZrS$_3$ & 0.22 & 1.13 \\
		$\alpha-$SrZrS$_3$ &0.22 & 6.35 \\
		$\beta-$SrZrS$_3$ &0.29 &0.61 \\
		BaZrS$_3$ & 0.19 & 0.33 \\
		\hline
	\end{tabular}
\end{table} 
where, $\textrm{R}_\infty$ is the Rydberg constant. Here, the $\varepsilon_{eff}$ lies in between the static value of dielectric constants as contributed by electrons and ions. The electronic and ionic static dielectric constants provide the upper and lower bounds to the exciton binding energy. For CaZrS$_3$, $\alpha-$SrZrS$_3$, $\beta-$SrZrS$_3$ and BaZrS$_3$, the electronic static dielectric constants are 4.06, 3.55, 3.44 and 4.19, respectively, which are calculated using BSE. The respective ionic static dielectric constants are 57.07, 20.86, 74.51, and 99.74 calculated using DFPT (see Figure \ref{fig_ionic}). For BaZrS$_3$, the calculated value of ionic static dielectric constant is in close agreement with previous experimental results\cite{PhysRevMaterials.4.091601}. From the reduced mass (provided in Table \ref{tbl:2}), and the static dielectric constants, we have determined the upper and lower bounds of E$_\textrm{B}$ using Equation~\ref{eq}, which are listed in Table \ref{tbl:bound} (also shown in Figure S2). The upper bounds are in good agreement with the E$_\textrm{B}$ calculated by taking the difference of GW and BSE peak positions, except for $\alpha-$SrZrS$_3$. Thus, the electronic contribution is more prominent than ionic contribution in dielectric screening for chalcogenide perovskites.

Further, we have determined the electron-phonon coupling using the Fr\"{o}hlich model~\cite{C6MH00275G,PhysRevB.96.195202}. In this model, the electron moving through the lattice interact with the polar optical phonons via Fr\"{o}hlich parameter $\alpha$, given by
\begin{equation}
\alpha=\left(\frac{1}{\upvarepsilon_\infty}-\frac{1}{\upvarepsilon_\textrm{static}}\right)\sqrt{\frac{\textrm{R}_\infty}{ch\omega_\textrm{LO}}}\sqrt{\frac{\textrm{m}^*}{\textrm{m}_\textrm{e}}}
\label{eq2}
\end{equation}
where $\upvarepsilon_\infty$ and $\upvarepsilon_\textrm{static}$ are the electronic and ionic static dielectric constants, respectively. $h$ is Planck's constant and $c$ is the speed of light. The characteristic frequency $\omega_\textrm{LO}$ is determined from the multiple phonon branches using athermal `B' scheme of Hellwarth \textit{et al}~\cite{PhysRevB.60.299}. The calculated values of $\alpha$ are provided in Table~\ref{tbl:ep}. From these, the reduction in QP gap can be determined~\cite{bokdam2016role}. For CaZrS$_3$, $\alpha-$SrZrS$_3$, $\beta-$SrZrS$_3$ and BaZrS$_3$, the QP gap is lowered by 0.24, 0.23, 0.33 and 0.27 eV, respectively. On comparing these values with E$_\textrm{B}$, we infer that except for $\alpha-$SrZrS$_3$, the charge-separated polaronic state is more stable than the bound exciton. Furthermore, the computed polaron mobilities are given in Table S3 (see section IX in SI).  
\begin{table}
	\caption{Electron-phonon coupling parameters for chalcogenide perovskites}
	\label{tbl:ep}
	\centering
	\setlength\extrarowheight{+4pt}
	\begin{tabular}[c]{cccc}
		\hline
		Configurations &$\omega_\textrm{LO}$ (cm$^{-1}$)&$\alpha_\textrm{e}$& $\alpha_\textrm{h}$ \\
		\hline
		CaZrS$_3$ &142.03& 4.51 & 8.29 \\
		$\alpha-$SrZrS$_3$ &152.74& 3.56 & 8.49 \\
		$\beta-$SrZrS$_3$ &117.86& 5.61 & 11.1 \\
		BaZrS$_3$ &107.51& 4.68 & 9.53 \\
		\hline
	\end{tabular}
\end{table}

\begin{figure}[h]
	\centering
	\includegraphics[width=0.4\textwidth]{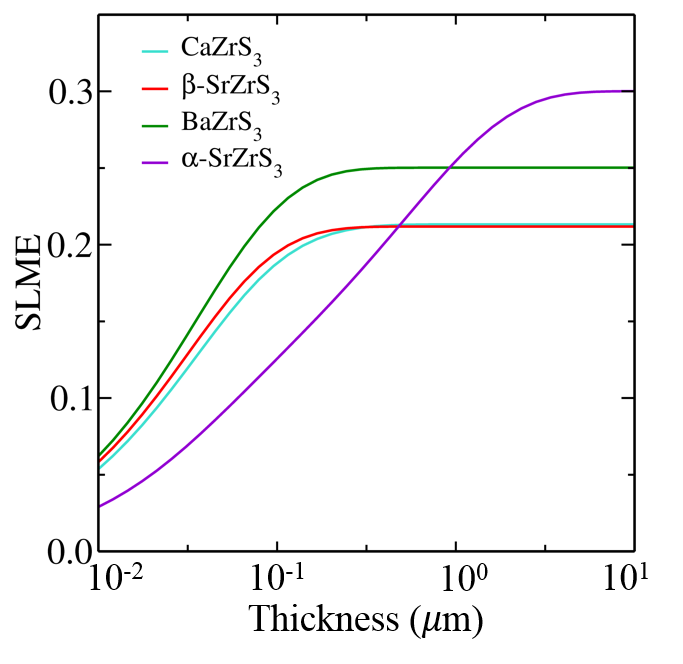}
	\caption{Spectroscopic limited maximum efficiency of AZrS$_3$ (A=Ca, Sr, and Ba).}
	\label{fig_slme}
\end{figure}

We have observed that, the AZrS$_3$ (A=Ca, Sr, Ba) perovskites exhibit large absorption coefficient and direct band gap in visible region. These two make them interesting materials for photovoltaic applications. Therefore, we have calculated the spectroscopic limited maximum efficiency (SLME)~\cite{PhysRevLett.108.068701,slme}, which has been proved to be a good metric to determine the maximum efficiency that an absorber material can reach in a single-junction solar cell. SLME is an improved version of Shockley and Queisser (SQ) efficiency,~\cite{doi:10.1063/1.1736034} as it takes into account the nature of band gap, the shape of absorption spectra and the material-dependent nonradiative recombination losses, in addition to the band gap. The standard solar spectrum, material's band gap and the absorption coefficient are given as input for the SLME calculation. Figure \ref{fig_slme} shows the calculated SLME of AZrS$_3$ (A=Ca, Sr, Ba).
Except for $\alpha-$SrZrS$_3$, the SLME becomes constant for a layer thickness greater than 1 $\mu$m. $\alpha-$SrZrS$_3$ requires a thicker absorption layer ($\sim$10 $\mu$m) for maximum efficiency. This is due to the fact that absorption onset in case of $\alpha-$SrZrS$_3$ is at larger value than the band gap. SLME values at 1 $\mu$m absorber layer thickness are 21.33\%, 25.45\%, 21.19\%, and 25.02\% for CaZrS$_3$, $\alpha-$SrZrS$_3$, $\beta-$SrZrS$_3$ and BaZrS$_3$, respectively (see Figure \ref{fig_slme}). The theoretically predicted SLME of BaZrS$_3$ (25.02\%) is in good agreement with previously reported theoretical efficiency ($\sim$25\% at 1 $\mu$m thickness)\cite{doi:10.1021/acs.chemmater.5b04213}. These values of SLME are encouraging for their photovoltaic applications.

In conclusion, we have determined the electronic and optical properties of chalcogenide perovskites AZrS$_3$ (A = Ca, Sr, Ba) by state-of-the-art ground- and excited-state methods. 
The effective mass of electron has been observed as decreasing down the group from Ca to Sr, thereby enhancing the charge carrier transport. 
Subsequently, the optical band gap is well reproduced by solving the Bethe-Salpeter equation (BSE). The exciton binding energies for CaZrS$_3$, $\alpha-$SrZrS$_3$, $\beta-$SrZrS$_3$ and BaZrS$_3$ are computed as 0.23, 0.54, 0.25 and 0.21 eV, respectively. In addition, by calculating the electron-phonon coupling parameters, we have observed that the charge-separated polaronic state is more stable than the bound
exciton. We also report negligible ionic contribution to the effective dielectric screening that determines the exciton binding energy.
Finally, the calculated spectroscopic limited maximum efficiency (SLME) suggests their usage in photovoltaics.

\section{Computational Methods}
The density functional theory (DFT)~\cite{PhysRev.136.B864,PhysRev.140.A1133} calculations have been performed as implemented in the Vienna \textit{ab initio} simulation package (VASP)~\cite{KRESSE199615,PhysRevB.59.1758}. The ion-electron interactions in all the elemental constituents are described using projector-augmented wave (PAW) potentials~\cite{PhysRevB.50.17953,PhysRevB.59.1758}. All the structures are optimized using generalized gradient approximation viz. PBE~\cite{PhysRevLett.77.3865}  exchange-correlation ($\epsilon_\textrm{xc}$) functional until the forces are smaller than 0.001 eV/\AA. The electronic self consistency loop convergence is set to 0.001 meV, and the kinetic energy cutoff is set to 500 eV for plane wave basis set expansion. A $k$-grid of $7\times7\times5$ is used for Brillouin zone integration, which is generated using Monkhorst-Pack~\cite{PhysRevB.13.5188} scheme. The effective mass has been calculated by SUMO~\cite{ganose2018sumo} using parabolic fitting of the band edges. Advanced hybrid $\epsilon_\textrm{xc}$ functional HSE06~\cite{doi:10.1063/1.2404663} is used for the better estimation of band gap. Note that spin-orbit coupling (SOC) has not been taken into account because it negligibly affects the electronic structure of considered chalcogenide perovskites (see Table S1). In order to determine optical properties, Bethe-Salpeter equation (BSE)~\cite{PhysRevLett.80.4510,PhysRevLett.81.2312} has been solved on top of single shot GW~\cite{PhysRev.139.A796,PhysRevLett.55.1418} (G$_0$W$_0$) calculations. The initial step for G$_0$W$_0$ calculation is performed by the PBE $\epsilon_\textrm{xc}$ functional. The polarizability calculations are carried out on a grid of 50 frequency points. The number of unoccupied bands is set to eight times the number of occupied orbitals (for convergence of empty states, see Table S2). $\Gamma$-centered $3\times3\times2$ $k$-grid has been used, for BSE calculations. To construct the electron-hole kernel for BSE calculations, 24 occupied and 24 unoccupied states have been used. The convergence for the same is shown in Figure S3. To check the convergence with respect to Brillouin zone sampling, model-BSE (mBSE) has been done (see Figure S4). We have found that there is a negligible shift of the lower energy peak on increasing the $k$-grid. Further, we have seen that the chalcogenide perovskites are optically active along all the three directions, signifying minute anisotropy of dielectric function in chalcogenide perovskites (see Figure S5). The ionic contribution to dielectric function has been calculated using density functional perturbation theory (DFPT) with $7\times7\times5$ $k$-grid generated using Monkhorst-Pack scheme.
\begin{acknowledgement}
MK acknowledges CSIR, India, for the senior research fellowship [grant no. 09/086(1292)/2017-EMR-I]. AS acknowledges IIT Delhi for the financial support. DG acknowledges UGC, India, for the senior research fellowship [grant no. 1268/(CSIR-UGC NET JUNE 2018)].
SB acknowledges the financial support from SERB under core research grant (grant no. CRG/2019/000647). We acknowledge the High Performance Computing (HPC) facility at IIT Delhi for computational resources.
\end{acknowledgement}
\begin{suppinfo}
Electronic band structure of chalcogenide perovskites using PBE $\epsilon_\textrm{xc}$ functional, exciton binding energy (E$_\textrm{B}$) of chalcogenide perovskites as a function of dielectric constant, effect of spin-orbit coupling (SOC) on band gap, convergence of number of bands used in G$_0$W$_0$ calculations, convergence of number of valence (NO) and conduction bands (NV) used in electron-hole interaction kernel for BSE calculations, imaginary part of dielectric function for BaZrS$_3$ with different number of $k$-grid using model-BSE (mBSE), imaginary part of dielectric function for BaZrS$_3$ with light polarization along the three lattice vectors, relations of excitonic parameters with exciton binding energy (E$_\textrm{B}$), dielectric constant ($\upvarepsilon_\textrm{eff}$) and reduced mass ($\mu$), polaron mobility of chalcogenide perovskites
\end{suppinfo}
\bibliography{achemso-demo}
\end{document}


\begin{center}
{\Large \bf Supplemental Material}\\ 
\end{center}
\begin{enumerate}[\bf I.]
	\item Electronic band structure of chalcogenide perovskites using PBE $\epsilon_\textrm{xc}$ functional
	\item Exciton binding energy (E$_\textrm{B}$) of chalcogenide perovskites as a function of dielectric constant
	\item Effect of spin-orbit coupling (SOC) on band gap
	\item Convergence of number of bands used in G$_0$W$_0$ calculations
	\item Convergence of number of valence (NO) and conduction bands (NV) used in electron-hole interaction kernel for BSE calculations
	\item Imaginary part of dielectric function for BaZrS$_3$ with different number of $k$-grid using model-BSE (mBSE)
	\item Imaginary part of electronic dielectric function for BaZrS$_3$ with light polarization along the three lattice vectors
	\item Relations of excitonic parameters with exciton binding energy (E$_\textrm{B}$), dielectric constant ($\upvarepsilon_\textrm{eff}$) and reduced mass ($\mu$)
	\item Polaron mobility of chalcogenide perovskites
\end{enumerate}
\vspace*{12pt}
\clearpage
\section{Electronic band structure of chalcogenide perovskites using PBE $\epsilon_\textrm{xc}$ functional}
\begin{figure}[h]
	\centering
	\includegraphics[width=0.8\textwidth]{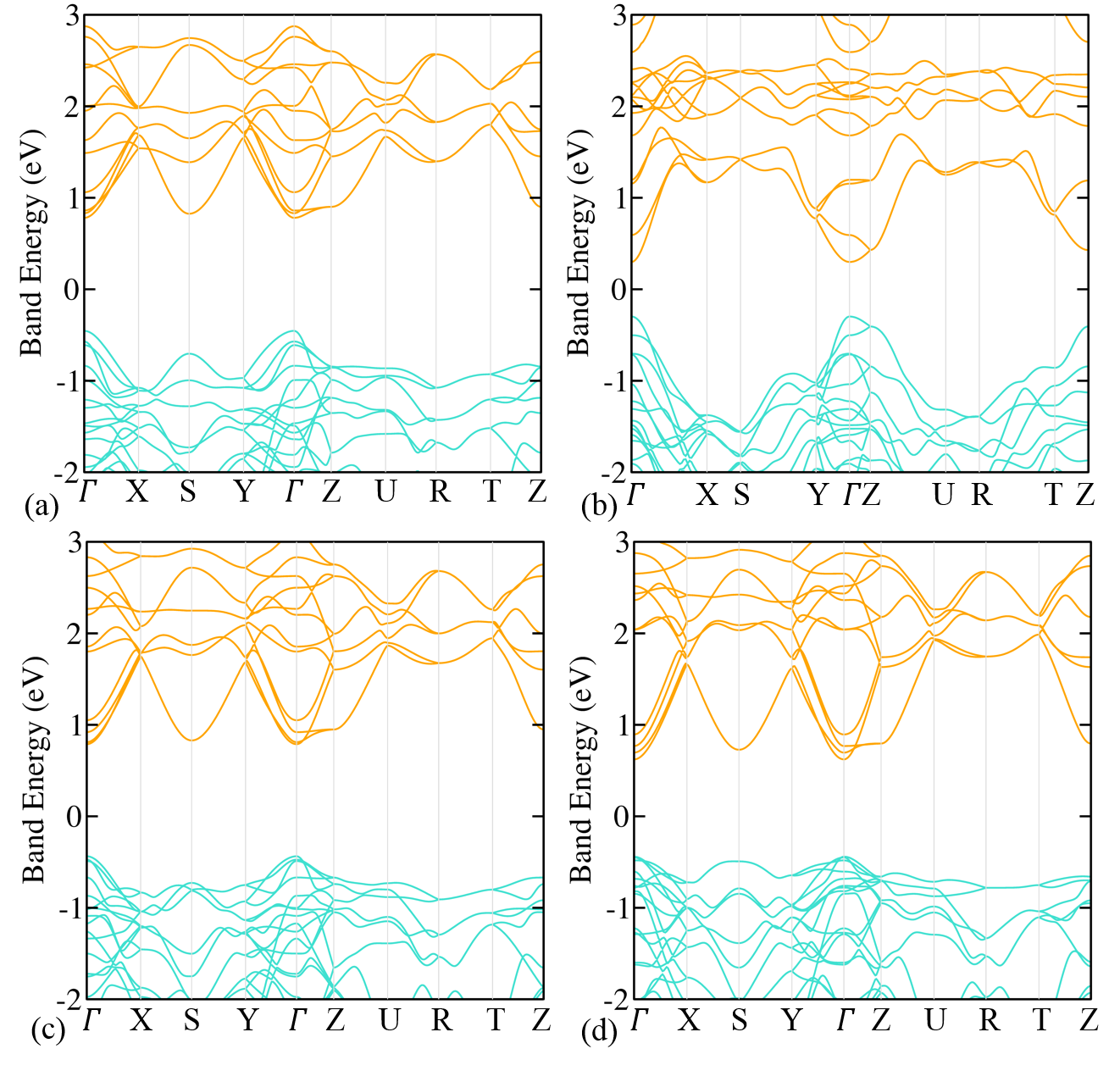}
	\caption{Electronic band structure of (a) CaZrS$_3$, (b) $\alpha-$SrZrS$_3$, (c) $\beta-$SrZrS$_3$, and (d) BaZrS$_3$ using PBE $\epsilon_\textrm{xc}$ functional.}
	\label{fig_band}
\end{figure}
\newpage
\section{Exciton binding energy (E$_\textrm{B}$) of chalcogenide perovskites as a function of dielectric constant}
\begin{figure}[h]
	\centering
	\includegraphics[width=0.8\textwidth]{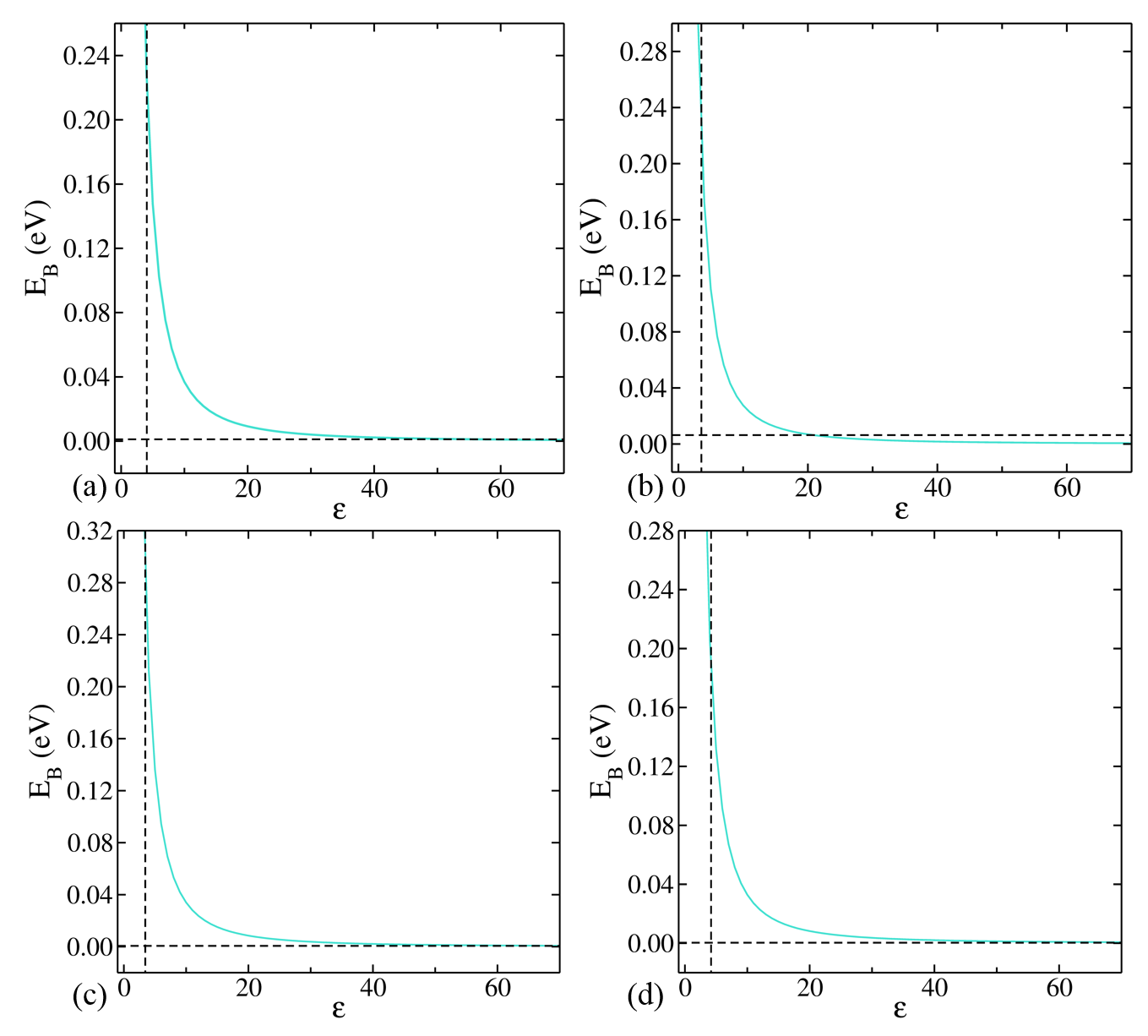}
	\caption{Exciton binding energy (E$_\textrm{B}$) of (a) CaZrS$_3$, (b) $\alpha-$SrZrS$_3$, (c) $\beta-$SrZrS$_3$, and (d) BaZrS$_3$, as a function of dielectric constant. The intersection of the curve with vertical dashed line defines the upper bound obtained using electronic static dielectric function and horizontal dashed line defines the lower bound obtained by ionic static dielectric function.}
	\label{fig_eps}
\end{figure}
\newpage
\section{Effect of spin-orbit coupling (SOC) on band gap}
\begin{table}
	\caption{Band gap (in eV) of CaZrS$_3$, $\alpha-$SrZrS$_3$, $\beta-$SrZrS$_3$, and BaZrS$_3$ using PBE $\epsilon_\textrm{xc}$ functional}
	\label{tbl:1}
	\centering
	\setlength\extrarowheight{+4pt}
	\begin{tabular}{p{3.8cm}p{2.6cm}p{2.0cm}}
		\hline
		Configurations & Without SOC & With SOC\\
		\hline
		CaZrS$_3$ & 1.24& 1.22\\
		$\alpha-$SrZrS$_3$ &0.60& 0.59\\
		$\beta-$SrZrS$_3$ &1.22&1.19\\
		BaZrS$_3$ &1.06&1.01\\
		\hline
	\end{tabular}
\end{table}
\newpage
\section{Convergence of number of bands used in G$_0$W$_0$ calculations}
\begin{table}
	\caption{Band gap (in eV) of CaZrS$_3$, $\alpha-$SrZrS$_3$, $\beta-$SrZrS$_3$, and BaZrS$_3$ using G$_0$W$_0$@PBE with different number of bands}
	\label{tbl:2}
	\centering
	\setlength\extrarowheight{+4pt}
	\begin{tabular}{p{3.8cm}p{2.4cm}p{2.4cm}p{2.4cm}p{2.4cm}}
		\hline
		Number of bands & CaZrS$_3$ &$\alpha-$SrZrS$_3$&  $\beta-$SrZrS$_3$& BaZrS$_3$\\
		\hline
		 320& 2.39& 1.63&2.42&2.20\\
		480&2.32& 1.60&2.35&2.12\\
		640&2.29&1.60&2.32&2.10\\
		720&2.29&1.60&2.32&2.10\\
		800&2.30&1.61&2.32&2.09\\
		\hline
	\end{tabular}
\end{table}
The number of bands used in the calculations are 720.
\newpage
\section{Convergence of number of valence (NO) and conduction bands (NV) used in electron-hole interaction kernel for BSE calculations}
\begin{figure}[h]
	\centering
	\includegraphics[width=0.6\textwidth]{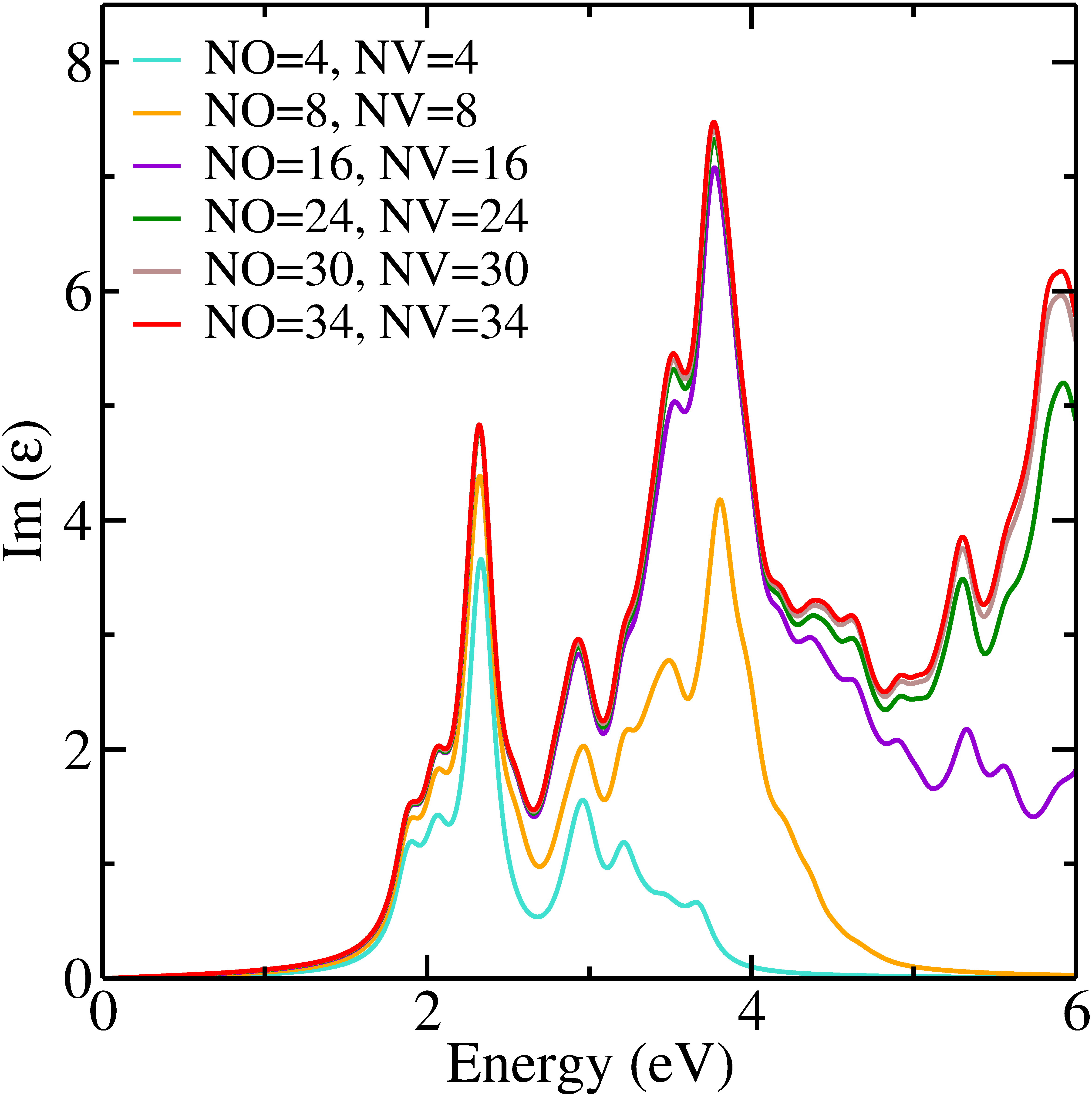}
	\caption{Imaginary part of electronic dielectric function with light polarization perpendicular to c-axis ($\upvarepsilon_\textrm{xx}$) for BaZrS$_3$ with different number of valence (NO) and conduction bands (NV) used in electron-hole interaction kernel.}
	\label{fig_nonv}
\end{figure}
\newpage
\section{Imaginary part of dielectric function for BaZrS$_3$ with different number of $k$-grid using model-BSE (mBSE)}
In model-BSE (mBSE), the dielectric function is replaced by the model local function given below:
\begin{equation}
\upvarepsilon^{-1}_\textrm{G,G}(\textrm{q})=1-(1-\upvarepsilon_\infty^{-1})\textrm{exp}\left(-\frac{|\textrm{q}+\textrm{G}|^{2}}{4\lambda^2}\right)
\end{equation}
where $\upvarepsilon_\infty$ is the static ion-clamped dielectric function in high frequency limit calculated using G$_0$W$_0$@PBE. $\lambda$ is the screening length parameter, determined by fitting the $\upvarepsilon^{-1}$ at small wave vector (q) with respect to $|\textrm{q}+\textrm{G}|$ and G is the reciprocal lattice vector.
\begin{figure}[h]
	\centering
	\includegraphics[width=0.9\textwidth]{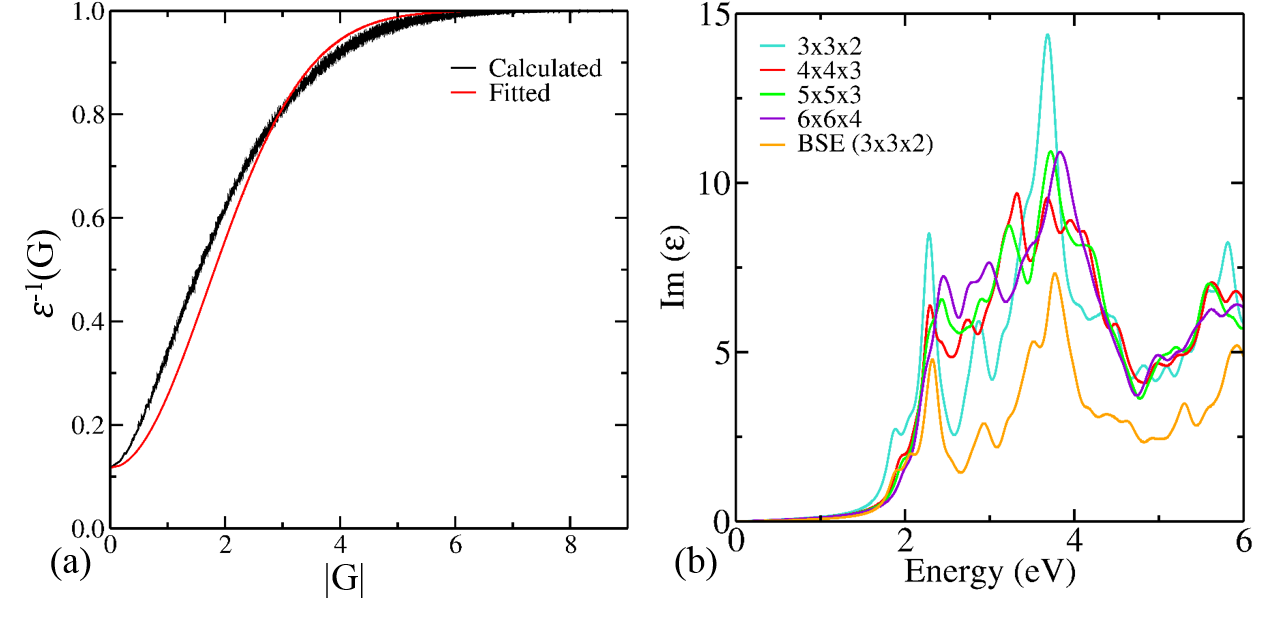}
	\caption{(a) Model fitting for model-BSE (mBSE). (b) Spatially average imaginary [Im $(\upvarepsilon)$] part of the dielectric function for BaZrS$_3$ with different $k$-grid using mBSE. Imaginary part using GW-BSE is shown for reference by orange color. Calculated values of inverse of the static ion-clamped dielectric function $\upvarepsilon_\infty^{-1}$ = 0.117 and the screening length parameter $\lambda$ = 1.20 are used in mBSE.}
	\label{fig_mbse}
\end{figure}
\newpage
\section{Imaginary part of electronic dielectric function for BaZrS$_3$ with light polarization along the three lattice vectors}
\begin{figure}[h]
	\centering
	\includegraphics[width=0.6\textwidth]{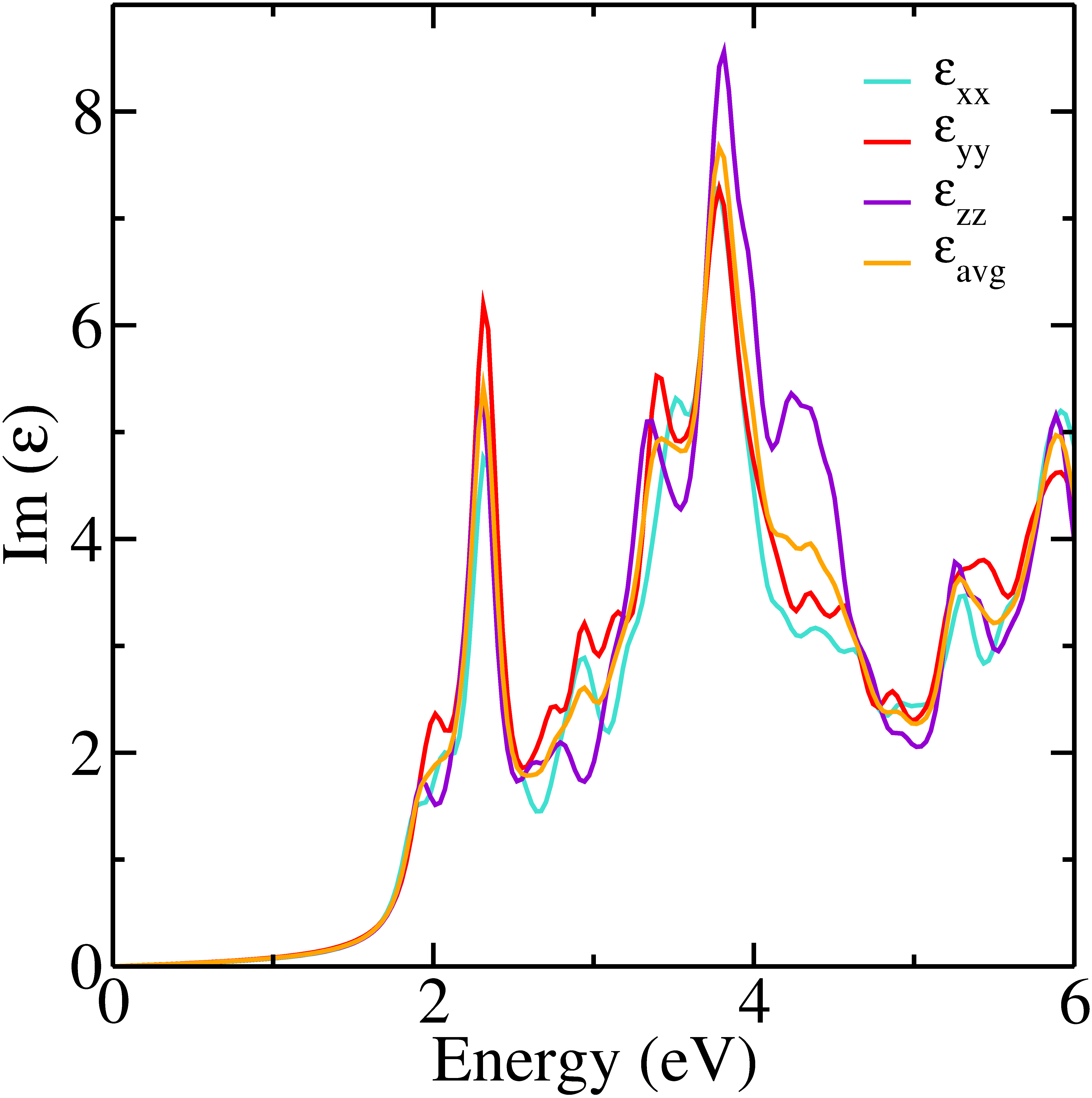}
	\caption{Imaginary part of electronic dielectric function for BaZrS$_3$ with light polarization along all three lattice vectors. For other chalcogenide perovskites as well, the anisotropy in dielectric function is negligible.}
	\label{fig_iso}
\end{figure}
\newpage
\section{Relations of excitonic parameters with exciton binding energy (E$_\textrm{B}$), dielectric constant ($\upvarepsilon_\textrm{eff}$), band gap (E$_\textrm{g}$) and reduced mass ($\mu$)}
The exciton temperature (T$_\textrm{exc}$) is determined as follows:
\begin{equation}
\textrm{T}_\textrm{exc}=\frac{\textrm{E}_\textrm{B}}{\textrm{k}_\textrm{B}}
\end{equation}
where E$_\textrm{B}$ is the exciton binding energy and $\textrm{k}_\textrm{B}$ is the Boltzmann constant. The exciton radius ($\textrm{r}_\textrm{exc}$) is calculated as follows:
\begin{equation}
\textrm{r}_\textrm{exc}=\frac{\textrm{m}_0}{\mu}\upvarepsilon_\textrm{eff}\,\textrm{n}^2\textrm{r}_\textrm{Ry}
\end{equation}
where $\textrm{m}_0$ is the free electron mass, $\mu$ is the reduced mass, $\upvarepsilon_\textrm{eff}$ is the static effective dielectric constant (here, the electronic dielectric constant has been taken since the ionic contribution to dielectric screening is negligible), n is the exciton energy level (n=1 provides the smallest exciton radius) and $\textrm{r}_\textrm{Ry}=0.0529$ nm is the Bohr radius. The excitonic lifetime ($\tau$) is inversely proportional to the probability of wavefunction for electron-hole pair at zero seperation ($|\phi_\textrm{n}(0)|^2$). 
The ($|\phi_\textrm{n}(0)|^2$) is determined as follows:
\begin{equation}
|\phi_\textrm{n}(0)|^2=\frac{1}{\pi(\textrm{r}_\textrm{exc})^3\textrm{n}^3}
\end{equation}
\newpage
\section{Polaron mobility of chalcogenide perovskites}
At temperature $T$, the Hellwarth polaron mobility can be determined from the static dielectric constants, effective mass and optical phonon frequency~\cite{PhysRevB.96.195202}.
\begin{table}
	\caption{Polaron mobilities ($\mu$) of CaZrS$_3$, $\alpha-$SrZrS$_3$, $\beta-$SrZrS$_3$, and BaZrS$_3$ at $T=300$ K}
	\label{tbl:3}
	\centering
	\setlength\extrarowheight{+4pt}
	\begin{tabular}{ccc}
		\hline
		Configurations & $\mu_\textrm{e}$ (cm$^2$/Vs) & $\mu_\textrm{h}$ (cm$^2$/Vs)\\
		\hline
		CaZrS$_3$ &8.16 & 5.96\\
		$\alpha-$SrZrS$_3$ &18.77& 6.81\\
		$\beta-$SrZrS$_3$ &6.84&3.76\\
		BaZrS$_3$ &11.35&5.58\\
		\hline
	\end{tabular}
\end{table}
{}